\documentclass[pra,final,twocolumn,showpacs,showkeys]{revtex4}
\usepackage{amsmath}
\usepackage{graphicx}
\usepackage{amsfonts}
\usepackage{amssymb}
\usepackage{subfigure}

\def\identity{\leavevmode\hbox{\small1\kern-3.2pt\normalsize1}}

\begin{document}

\title{Reexamination of decoherence in quantum walks on the hypercube}

\author{Frederick W. Strauch}
\email[Electronic address: ]{Frederick.W.Strauch@williams.edu}
\affiliation{Department of Physics, Williams College, Williamstown, MA 01267}

\date{\today}

\begin{abstract}
The effect of decoherence on the continuous-time quantum walk on the hypercube is revisited.  Previously, an exact solution was found for a decoherence model that preserved the effective tensor-product form of the dynamics.  Here a new model is presented, inspired by perfect quantum state transfer in qubit networks.  A perturbative solution is found for the dynamics of this model which is not of a tensor-product form.  In contrast to previous results, the hitting probability has a lower bound that is independent of the hypercube dimension.  
\end{abstract} 
\maketitle

\section{Introduction}

The quantum walk on the hypercube is a quintessential problem in quantum information processing.  The main uses of quantum walks can be divided into three parts: search, hitting, and fast sampling algorithms.  The first, the search of a database, is based on Grover's algorithm, and can be implemented using a quantum walk on a hypercube graph \cite{Shenvi2003}.  Hitting algorithms are those in which a particle can propagate from one node to another node in a graph (sometimes called the graph traversal problem).  For the hypercube, it has been shown that the quantum walk performs this task exponentially faster than classical random walks \cite{Kempe2003a}; a graph for which the quantum walk is exponentially faster than all classical algorithms was found by Childs {\it et al.} \cite{Childs2003}.  Finally, a fast sampling algorithm approximates a probability distribution (often uniform) over some graph starting from a simple initial condition---this is quantified by what is called the mixing time.  The hitting and instantaneous mixing times for the hypercube were found in the seminal work by Moore and Russell \cite{Moore2001}.   Other important works on the hypercube include recent results on the hitting  \cite{Krovi2006} and mixing times \cite{Abal2008}; there have also been general reviews on quantum walks \cite{Kempe2003,Kendon2006}.

In addition to potential algorithms, the quantum walk on the hypercube can be used to devise qubit networks that perfectly transfer a quantum state between nodes \cite{Christandl2004}, and could possibly be implemented using superconducting qubits \cite{Strauch2007}.  For these physical implementations, one must consider the inevitable coupling of each qubit to its environment.  The simplest such model leads to a particular decoherence model, the solution of which is the subject of this paper.  More generally, decoherence in quantum walks may be useful \cite{Kendon2003,Kendon2006}, and could even have implications for biological systems \cite{Mohseni2008}.

Previous work on decoherence in quantum walks on the $d$-dimensional hypercube began with the work by Kendon and Tregenna \cite{Kendon2003}, who looked at a discrete-time quantum walk interrupted by measurements.  They found numerically that by measuring the position of the walk with probability $p$ after each time step the first hitting probability was exponentially (in $d$) suppressed.  An exact solution for decoherence in a continuous-time walk was found by Alagi\'{c} and Russell \cite{Alagic2005}, using a model where dephasing occurred between different subspaces of the hypercube.  Here too, the first hitting probability decays exponentially in $d$.  

In this paper I consider a continuous-time decoherence model---here called the vertex model---that is similar to the discrete-time model mentioned above \cite{Kendon2003}.  The specific form of this model arose from a study of decoherence in quantum state transfer \cite{Strauch2007}.  By analyzing the master equation of the vertex model perturbatively, an analytical solution is found, good for weak decoherence.  For this model and a natural choice of parameters the quantum probability to traverse a $d$-dimensional hypercube remains bounded for all $d$, in opposition to the exponential suppression found previously.  

This paper is organized as follows.  The decoherence models are presented in Section II along with a review of the exact results for the subspace model \cite{Alagic2005}.  In Section III, a perturbative approach is presented, giving an analytical expression for the hitting probability as a function of time.  In Section IV, I compare this model with the discrete and continuous models, and discuss the role of information lost due to decoherence.  Finally, I conclude in Section V by summarizing these results and their implications.  An Appendix extends the results of Ref. \cite{Strauch2007} and shows how different models of decoherence in a hypercube network of qubits generates the decoherence models considered here.

\section{Decoherence Models}
In this section I first describe two models of decoherence: the subspace model studied by Alagi\'{c} and Russell \cite{Alagic2005} and the vertex model presented here.  The quantum walk on the hypercube involves the continuous-time dynamics (with $\hbar = 1$)
\begin{equation}
|\Psi(t)\rangle = e^{-i H t} |\Psi(0)\rangle,
\end{equation} 
where the matrix elements of $H$ are proportional to the adjacency matrix of the $d$-dimensional hypercube. The vertices of the hypercube are labeled by bit-strings of length $d$, and two vertices are connected if their labels differ by one bit.  This induces a tensor-product structure, in terms of which the Hamiltonian has the simple form
\begin{equation}
H = \omega \sum_{j=1}^d \identity \otimes \cdots \sigma_x \dots \otimes \identity,
\end{equation}
where the Pauli matrix $\sigma_x$ occurs at location $j$.  Here and in the following, $\omega$ and $t$ are taken to be dimensionless quantities.  Using this Hamiltonian, one has 
\begin{equation}
| \Psi(t) \rangle = ( e^{-i \omega t \sigma_x} )^{\otimes d} |\Psi(0)\rangle.
\end{equation}
Letting $|a\rangle = |0 \cdots 0 \rangle$ and $|b\rangle = |1 \cdots 1 \rangle$, the transfer probability from $|\Psi(0)\rangle = |a\rangle$ to $|b\rangle$ is given by 
\begin{equation}
P(t) = | \langle b | e^{-i H t} | a \rangle |^2 = | \sin(\omega t) |^{2 d}.
\end{equation}
The first hitting time occurs at $T = \pi / (2 \omega)$, with unit probability.

In the presence of decoherence, however, this simple dynamics is modified.  First, the system is no longer described by a wavefunction but rather by a density matrix.  Second,  the evolution is no longer generated by a Hamiltonian operator, but rather by a superoperator, here assumed to be of the general Lindblad form \cite{Breuer2002}
\begin{equation}
\partial_t \rho(t) = - i [ H, \rho(t)] + \sum_{j} \lambda_j  (L_j \rho L_j^{\dagger} - \frac{1}{2} L_j^{\dagger} L_j \rho - \frac{1}{2} \rho L_j^{\dagger} L_j ).
\end{equation}
The $L_j$ are called the Lindblad operators, with rates $\lambda_j$.  

Alagi\'{c} and Russell considered Lindblad operators of the form:
\begin{equation}
L_{j,\alpha} =  \identity \otimes \cdots \Pi_{\alpha} \cdots \otimes \identity
\end{equation}
where $\alpha$ is 0 or 1 (these should be included in the sum) and $\Pi_{0} = |0\rangle \langle 0 |$ and $\Pi_{1} = |1\rangle \langle 1|$ are the projectors for coordinate $j$.  The decoherence rates are all identical $\lambda_j = \lambda$ (see Section IV for a discussion of these rates).  Note that these operators do not project the state onto any particular node of the hypercube network, but rather onto one of two subspaces in which the node has a $0$ (or $1$) in the $j$-th bit.  For this reason, I will call this the subspace model.  

The resulting master equation can be given a tensor-product form, which allows for the solution of the density matrix by solving for the eigenvalues and eigenvectors of a $4 \times  4$ matrix \cite{Alagic2005}.  Using these and the initial condition $\rho(0) = |a\rangle \langle a|$, the hitting probability is
\begin{equation}
\begin{array}{lcl}
P_s(t) &=& \langle b | \rho(t) | b \rangle \\
&=& 2^{-d} \left[ 1- e^{-\lambda t/2}  \left\{\cos(\beta t/2) + \frac{\lambda}{\beta} \sin(\beta t/2)\right\}\right]^d
\end{array}
\label{ssprob}
\end{equation}
with $\beta = \sqrt{16 \omega^2 - \lambda^2}$.

The decoherence model considered here uses the same general form, again with identical rates  $\lambda_j = \lambda$ but with Lindblad operators of the form
\begin{equation}
L_{\alpha_1, \cdots, \alpha_d} =  \Pi_{\alpha_1} \otimes \cdots \otimes \Pi_{\alpha_d},
\end{equation}  
each $\alpha$ being 0 or 1.  These operators project a state onto the vertex given by $|\alpha_1 \cdots \alpha_d\rangle$.  I call this the vertex model, and note that it can be derived from the master equation for qubits undergoing dephasing independently in a hypercube network, as shown in the Appendix.  Note also that this decoherence model has essentially nothing in common with the graph on which the quantum walk occurs.  This is in constrast to the subspace model, in which decoherence somehow knows the structure of the network.  The vertex model is somewhat more physical, in the sense that for a physical network, decoherence is more likely to affect each vertex independently, and less likely to know about the full hypercube structure.  This will be discussed in Sec. IV.

It is instructive to reexpress the master equations in terms of the matrix elements $\rho_{x,y} = \langle x | \rho | y \rangle$, where $x = x_1 \cdots x_d$ and $x_j = \{0,1\}$.  For the vertex model, $\rho_{x,y} $ satisfies
\begin{equation}
\partial_t \rho_{x,y} = -i \sum_{z} \left(H_{x,z} \rho_{z,y} - \rho_{x,z} H_{z,y}\right) - \lambda (1-\delta_{x,y})\rho_{x,y},
\label{vmodel}
\end{equation}
where $\delta_{x,y}$ is the Kronecker delta and $H_{x,y} = \langle x |H| y \rangle$.  This master equation involves the decay of all off-diagonal matrix elements with rate $\lambda$, consistent with dephasing of every vertex.  The subspace model has
\begin{equation}
\partial_t \rho_{x,y} = -i \sum_{z} \left(H_{x,z} \rho_{z,y} - \rho_{x,z} H_{z,y}\right) - \lambda \sum_j (1-\delta_{x_j,y_j})\rho_{x,y},
\label{ssmodel}
\end{equation}
where $\delta_{x_j,y_j}$ is the Kronecker delta for the $j$-th bit of $x$ and $y$.  This too involves the decay of the off-diagonal elements, but these are now weighted by the number of bits in which the indices $x$ and $y$ differ.  Those elements with indices differing in $n$ bits decay with a rate $n \lambda$.   This is somewhat surprising---the subspace model presumably gets less information about the location of the ``walker'' in the quantum walk, yet in the end causes more decoherence than the vertex model.  This will be discussed in Section IV.

\section{Perturbative Solution}
In the limit of weak decoherence $\lambda \ll \omega$, perturbation theory can be used, starting from the eigenstates of $H$.  However, since this is a master equation, careful treatment of degeneracies is required.  These degeneracies could be analyzed using an angular momentum representation or perhaps other group theory methods.  Here, however, I will take a direct approach, requiring the following notation.  The density matrix of the hypercube can be rewritten as  a vector, using the basis states
\begin{equation}
|x_1 \cdots x_d ; y_1 \cdots y_d \rangle = |x_1 \cdots x_d \rangle \langle y_1 \cdots y_d |,
\end{equation}
where each $x_j$ and $y_j$ are either 0 or 1.  These states will also be written using an abbreviated notation $|x;y\rangle$.  In this basis, the superhamiltonian $\mathcal{H}$ acts as
\begin{equation}
\mathcal{H} |x_1 \cdots x_d ; y_1 \cdots y_d \rangle \sim -i [H, |x_1 \cdots x_d \rangle \langle y_1 \cdots y_d | ]
\end{equation}
with the value
\begin{equation}
\begin{array}{lcl}
\mathcal{H} |x;y\rangle &=&-i \sum_{j=1}^{d} |x_1 \cdots \bar{x}_j \cdots x_d ; y_1 \cdots y_d \rangle  \\
&&+i \sum_{j=1}^{d} |x_1 \cdots x_d ; y_1 \cdots \bar{y}_j \cdots  y_d \rangle
\end{array}
\end{equation}
where $\bar{x}  = 1-x$ is the bit-flip of $x$.  The Lindblad superoperator $\mathcal{L}_0$ is defined similarly, and is diagonal in this basis:
\begin{equation}
\mathcal{L}_0 |x ; y \rangle = -\lambda (1-\delta_{x,y}) |x; y \rangle.
\label{lbasis1}
\end{equation}

In addition to this basis, I also introduce the eigenstates of $\mathcal{H}$, which have the form
\begin{equation}
|x ;  y\rangle_x = 2^{-d} \sum_{x',y'}  (-1)^{x \cdot x' + y \cdot y'}  | x'; y' \rangle, 
\label{basis2}
\end{equation}
using the abbreviated notation for the states, with $x,y,x',y'$ representing bit strings of size $d$, and thus there are $2^{2d}$ terms in the sum.  Note that I have also used the bit-wise product
\begin{equation}
x \cdot x'  = \sum_{j=1}^d x_j {x'}_j.
\end{equation}
These states satisfy
\begin{equation}
\mathcal{H} | x; y \rangle_x = 2 i \omega \left( \sum_{j=1}^d [x_j - y_j] \right) | x; y \rangle_x.
\label{basish}
\end{equation}

While $\mathcal{H}$ is diagonal in this basis, $\mathcal{L}_0$ is not.  To find how it acts in this basis, I first use Eq. (\ref{lbasis1}) and the definition of $|x;y\rangle_x$ to find
\begin{equation}
\mathcal{L}_0 |x ; y \rangle_x = - \lambda |x; y\rangle_x + \lambda 2^{-d} \sum_{x'} (-1)^{x \cdot x' + y \cdot x'} |x' ; x'\rangle.
\end{equation}
Then, using the inverse of Eq. (\ref{basis2})
\begin{equation}
|x ; y \rangle = 2^{-d} \sum_{x',y'}  (-1)^{x \cdot x' + y \cdot y'}  | x'; y' \rangle_x
\end{equation}
and the fact that
\begin{equation}
2^{-d} \sum_{x'} (-1)^{x \cdot x' + y \cdot x' + x' \cdot x'' + x' \cdot y''} = \delta_{x \oplus x'', y \oplus y''},
\end{equation}
where $x \oplus x''$ is bit-wise addition, I find the result
\begin{equation}
\mathcal{L}_0 |x ; y \rangle_x = - \lambda |x; y\rangle_x + \lambda 2^{-d} \sum_{x'',y''} \delta_{x \oplus x'', y \oplus y''}  |x'' ; y''\rangle_x.
\end{equation}
Now, since this final sum is over all bit-strings, I can make the substitutions $x'' = x \oplus z$ and $y'' = y \oplus z'$ and sum over $z$ and $z'$.  These substitutions reduce the delta function to $\delta_{z,z'}$, with the final result
\begin{equation}
\mathcal{L}_0 |x ; y \rangle_x = - \lambda |x; y\rangle_x + \lambda 2^{-d} \sum_{z} |x \oplus z ; y \oplus z\rangle_x.
\label{lbasis2}
\end{equation}

If all of the states of $\mathcal{H}$ were nondegenerate, I could simply assert that the effect of decoherence on each state is given by $ \langle x ; y| \mathcal{L}_0 | x; y \rangle_x$.  However, there are many degeneracies, and so degenerate perturbation theory is required.  From Eq. (\ref{basish}) I find that there are $2 d+1$ subspaces, with eigenvalues $\epsilon_n = 2 i \omega (d-n)$, with $n = 0, \cdots,2 d$.  The degeneracy of subspace $n$ can be found to be $(2d)! / [ n! (2d-n)!]$.   In the following, I will consider $\mathcal{L}_0$ projected onto these subspaces (with the same symbol) and analyze how it mixes those states within each subspace.

Finding the correct form of the eigenstates in these degenerate subspaces requires some additional notation.  Consider, for example, the subspace with $n=2$.  In this subspace, there can be no more than $2$ bits of $y$ that are unity, and no more than $2$ bits of $x$ that are zero.  Letting those bits occur at positions $i$ and $j$ (with $i<j$), this can be achieved in one of four ways:
\begin{equation}
\begin{array}{lcl}
|ij; \emptyset \rangle_2 &=& |1 \cdots 1 0_i 1 \cdots 1 0_j 1 \cdots 1; 0 \cdots 0\rangle_x \\
|i;j\rangle_2 &=& |1 \cdots 1 0_i 1 \cdots 1; 0 \cdots 0 1_j 0 \cdots 0 \rangle_x \\
|j;i\rangle_2 &=& |1 \cdots 1 0_j 1 \cdots 1; 0 \cdots 0 1_i 0 \cdots 0 \rangle_x \\
|\emptyset; ij \rangle_2 &=& |1 \cdots 1; 0 \cdots 0 1_i 0 \cdots 0 1_j 0 \cdots 0 \rangle_x.
\end{array}
\label{basis3}
\end{equation} 
Note that the indices $i$ and $j$ label the zeros for $x$ and the ones for $y$; these can be interpreted as the locations of ``excitations.''  Using Eq. (\ref{lbasis2}),  when projected to the $n=2$ subspace the Lindblad operator has the following effect
\begin{equation}
\mathcal{L}_0 |v \rangle_2 = - \lambda |v \rangle_2 + \lambda 2^{-d} (|ij;\emptyset \rangle_2 + |i;j\rangle_2 + |j;i\rangle_2 + |\emptyset; ij\rangle_2 )
\end{equation}
where $|v\rangle_2$ is any of the states in Eq. (\ref{basis3}). These are for $i<j$.  For $i=j$, however, \begin{equation}
\mathcal{L}_0 |j;j\rangle_2 = - \lambda |j;j\rangle_2 + \lambda 2^{-d} \sum_{k} |k;k\rangle_2.
\end{equation}
Thus, for the states in this subspace the effect of $\mathcal{L}_0$ is to either mix excitations between the $x$ and $y$ indices, or to distribute ``paired'' excitations over all possible locations.

States with higher $n$ can be defined and the effect of $\mathcal{L}_0$ can be analyzed in an analogous fashion.  For example $|j_1 \cdots j_n ; \emptyset\rangle_n$ denotes a state $|x;y\rangle$ with $n$ zeros in $x$, located at positions $j_1, j_2, \cdots, j_n$, and no ones in $y$.  To represent the effect of $\mathcal{L}_0$, it is convenient to define the symbol $\mathcal{S}$ as a type of shift operator, transferring the last unpaired excitation in $x$ to $y$, such that
\begin{equation}
\begin{array}{lcl}
\mathcal{S} |j_1 \cdots j_n; \emptyset \rangle_n &=& |j_1 \cdots j_{n-1} ; j_n \rangle_n \\
\mathcal{S}^2 |j_1 \cdots j_n; \emptyset \rangle_n &=& |j_1 \cdots j_{n-2};  j_{n-1} j_n \rangle_n \\
\vdots &=& \vdots \\
\mathcal{S}^n |j_1 \cdots j_n; \emptyset \rangle_n &=& |\emptyset; j_1 \cdots j_n \rangle_n.
\end{array}
\end{equation}
In addition, I define $\mathcal{P}_{m,n}$ to represent a permutation of $j_n$ with $j_m$.  For example, 
\begin{equation}
\mathcal{P}_{m,n} |j_1 \cdots j_m \cdots j_{n-2}; j_{n-1} j_n \rangle_n = |j_1\cdots j_{n-2} j_n; j_m j_{n-1} \rangle_n,
\end{equation}
where I have reordered the labels to increasing order, with $m < n - 1$.  Using these two symbols, I now argue that a correct set of states that diagonalize $\mathcal{L}_0$ (when projected to the relevant subspace) is given by
\begin{equation}
|\phi_{j,\vec{s}} \rangle_n = 2^{-n/2} \left( \prod_{m=1}^n [1+s_m \mathcal{S}] \mathcal{P}_{m,n} \right) |j_1 \cdots j_n; \emptyset \rangle_n,
\label{phidef}
\end{equation}
where $j = \{ j_1, \cdots, j_n\}$, $\vec{s} = (s_1, \cdots, s_n)$ and $s_{m} = \pm 1$.  When this product is expanded, it produces a superposition of states with the indices $j$ distributed in all possible ways ($2^n$) over the $x$ and $y$ partitions, and with corresponding weights given by the products of $s_m$.   Note that the effect of $\mathcal{L}_0$ can be expressed quite simply using this basis.  Specifically, for $|v\rangle_n$ any state formed by shifting or permuting $|j_1 \cdots j_n ; \emptyset\rangle_n$:
\begin{equation}
\mathcal{L}_0 |v\rangle_n = - \lambda |v\rangle_n + \lambda 2^{-d+n/2} |\phi_{j,\vec{1}}\rangle_n,
\end{equation}
where $\vec{1} = (1,\cdots,1)$.  
In fact, combining this result with Eq. (\ref{phidef}), and using the identity
\begin{equation}
\prod_{m=1}^n (1+s_m) = 2^n \delta_{\vec{s},\vec{1}},
\end{equation}
the following result is obtained:
\begin{equation}
\mathcal{L}_0 |\phi_{j,\vec{s}} \rangle_n = - \lambda (1 - 2^{n-d} \delta_{\vec{s},\vec{1}}) |\phi_{j,\vec{s}} \rangle_n.
\end{equation}

This enumerates the states with $j_1 < j_2 < \dots < j_n$.  For cases such as $j_{n-1} = j_n$, where two labels are paired (one with an excitation at site $j_n$ for $x$, the other for $y$), the states have a slightly different form.  With $p$ pairs in subspace $n$, the pairs can be distributed in $d_p = (d-n+2p)!/[p! (d-n+p)!]$ different ways.  Here I choose to use a discrete Fourier transform to construct states for these pairs, with the form 
\begin{widetext}
\begin{equation}
|\phi_{j,\vec{s},q} \rangle_{n,p} = \mathcal{N}_{n,p} \left( \prod_{m=1}^{n-2p} [1+s_m \mathcal{S}] \mathcal{P}_{m,n-2p} \right) \sum_{\{k\}} e^{i 2 \pi q f(k)/d_p } |j_1 \cdots j_{n-2p} k_1 \cdots k_p; k_1 \cdots k_p \rangle_n,
\end{equation}
\end{widetext}
where here $\mathcal{S}$ and $\mathcal{P}$ act only on the unpaired indices and the sum is over all integers with $1 \le k_1< k_2 < \cdots < k_p \le d$ and each $k_m \ne \{j_1, \cdots, j_{n-2p} \}$.  The function $f(k)$ labels each configuration of $k = \{k_1, \cdots, k_n\}$ from $0$ to $d_p-1$ (its explicit form is not needed here), and $0<q<d_p-1$.  The normalization factor is
\begin{equation}
\mathcal{N}_{n,p} = 2^{-n/2+p} d_p^{-1/2} = 2^{-n/2+p} \left(\frac{p! (d-n+p)!}{(d-n+2p)!}\right)^{1/2}.
\end{equation}
Note that for $p$ pairs there are $d!/[(n-2p)! (d-n+2p)!]$ ways of choosing the index set $j$, $2^{n-2p}$ possible choices of $\vec{s}$, and $d_p$ possible values of $q$.  Thus, the total number of these states is (for $n<d$)
\begin{equation}
\sum_{p=0}^{\lfloor n/2 \rfloor} \frac{d!}{(n-2p)! (d-n+2p)!} d_p 2^{n-2p} = \frac{(2d)!}{n! (2d-n)!},
\end{equation}
which---as required---equals the total degeneracy of subspace $n$.  Using an argument similar to the case of $p=0$ given above, these states can be shown to satisfy
\begin{equation}
\mathcal{L}_0 |\phi_{j,\vec{s},q} \rangle_{n,p} = - \lambda \left(1 - 2^{n-d-2p} d_p  \delta_{\vec{s},\vec{1}} \delta_{q,0} \right) |\phi_{j,\vec{s},q}\rangle_{n,p}.
\end{equation}

This completes the analysis of the effect of $\mathcal{L}_0$ on each subspace.  To calculate the hitting probability starting from the initial state $|a\rangle \langle a|$ and ending at $| b\rangle \langle b|$, I first expand these in terms of the eigenstates of $\mathcal{L}_0$ on each $(n,p)$ subspace:
\begin{equation}
c_{j,\vec{s},q} = {}_{n,p} \langle \phi_{j,\vec{s},q} | a;a\rangle = 2^{n/2-p-d} d_p^{1/2} \delta_{\vec{s},\vec{1}} \delta_{q,0}
\end{equation}
and
\begin{equation}
d_{j,\vec{s},q} = \langle b;b | \phi_{j,\vec{s},q} \rangle_{n,p} = (-1)^{d-n} c_{j,\vec{s},q}.
\end{equation}
Denoting the eigenvalues of $|\phi_{j,\vec{s},q}\rangle_{n,p}$ by $2i\omega(d-n) - \lambda_{pn}$ with
\begin{equation}
\lambda_{pn} = \lambda\left(1 - 2^{n-d-2p} \frac{(d-n+2p)!}{p! (d-n+p)!}\right),
\label{vlamb}
\end{equation}
the net result for the probability is
\begin{equation}
\begin{array}{lcl}
P_v(t) &=& \langle b; b | e^{(\mathcal{H} + \mathcal{L}_0)t} | a; a \rangle \\
&=& \sum_{n=0}^{2d} \sum_{p=p_{min}}^{\lfloor n/2 \rfloor} \sum_{j,\vec{s},q} d_{j,\vec{s},q} c_{j,\vec{s},q} e^{2 i \omega (d-n) t} e^{-\lambda_{pn} t},
\end{array}
\end{equation}
where $p_{min} = \max(0,n-d)$.  Substituting for $c_{j,\vec{s},q}$ and $d_{j,\vec{s},q}$ and performing the sum over $j$, $\vec{s}$, and $q$ leads to the result
\begin{equation}
P_v(t) = \sum_{n=0}^{2d} \sum_{p=p_{min}}^{\lfloor n/2 \rfloor} \frac{d! 2^{n-2p-2d} (-1)^{d-n}}{p! (n-2p)! (d-n+p)! } e^{2 i \omega (d-n) t} e^{-\lambda_{p n} t}.
\end{equation}
This can be simplified to some degree, by noting there is a symmetry between terms with $n< d$ and $n>d$ (this can be found explicitly by letting $n' = 2d - n$ and $p' = p + d - n$ for $n>d$).  Doing leads to the the final result
\begin{equation} 
P_v(t) = \sum_{n=0}^d (-1)^{d-n} g_n(t) \cos \left(2 \omega t (d-n) \right)
\label{vprob1}
\end{equation}
where I have defined the functions
\begin{equation}
g_n(t) = \sum_{p=0}^{\lfloor n/2 \rfloor} \frac{d! (2 - \delta_{n,d})2^{n-2p-2d}}{p! (n-2p)! (d-n+p)!} e^{-\lambda_{pn}t}.
\label{vprob2}
\end{equation}

\section{Comparison of Models}
The subspace and vertex models initially seem somewhat similar.  Each can be written in terms of functions that recall damped harmonic oscillators, and indeed, they agree for $\lambda \ll \omega$ and for $d=1$:
\begin{equation}
P_s(t,d=1) \approx P_v(t,d=1) = \frac{1}{2} \left(1 - e^{-\lambda t/2} \cos(2 \omega t)\right).
\end{equation}
For larger values of $d$, however, significant differences emerge.  

\begin{figure}
\begin{center}
\includegraphics[width = 3 in]{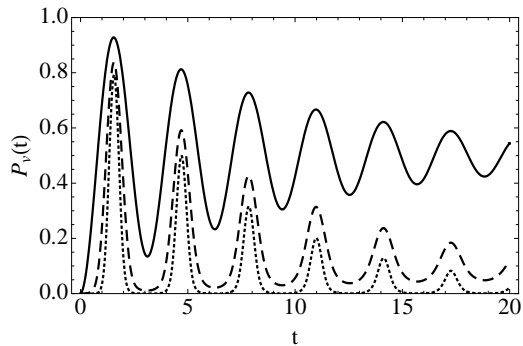}
\caption{The hitting probability $P_v(t)$ as a function of time for the quantum walk with decoherence in the vertex model.  The parameters are $\omega=1$, $\lambda = 1/5$, and $d=1$ (solid), $d=4$ (dashed), and $d=10$ (dotted). }
\label{qwvertex}
\end{center}
\end{figure}

\begin{figure}
\begin{center}
\includegraphics[width = 3 in]{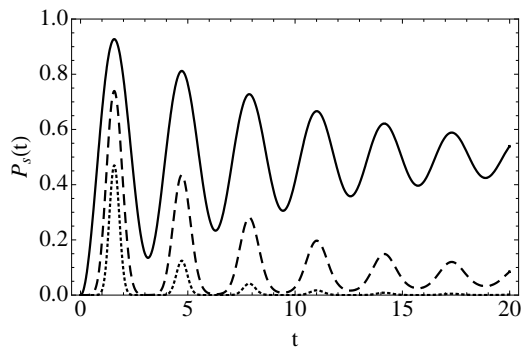}
\caption{The hitting probability $P_s(t)$ as a function of time for the quantum walk with decoherence in the subspace model.  The parameters are $\omega=1$, $\lambda = 1/5$, and $d=1$ (solid), $d=4$ (dashed), and $d=10$ (dotted). }
\label{qwsubspace}
\end{center}
\end{figure}

For illustration, I compare the two models with $\omega = 1$, $\lambda = 1/5$, and for $d$ equal to 1, 4, and 10.  In Fig.~\ref{qwvertex}, the hitting probability $P_v(t)$ is shown for the vertex model, using the perturbative solution presented in the previous section.  Note that this value of $\lambda$ was chosen so that the probabilities decayed sufficiently at the transfer time $T = \pi/(2\omega) = \pi/2$; here $\lambda T = \pi/10 \approx 0.314$.  Direct numerical simulations, using a split-operator algorithm \cite{McLachlan2002} to evolve the density matrix by Eq. (\ref{vmodel}) show that for these values of $\lambda$ and $d$, the perturbative and numerical solutions are in good agreement.  The hitting probability for the subspace model $P_s(t)$ is shown in Fig.~\ref{qwsubspace}.  There are several common features found in these two figures.  First, there is clearly damped oscillations of the probability, with a frequency approximately given by $2 \omega$.  Second, for long times the probability becomes constant and equal to $2^{-d}$.  

\begin{figure}
\begin{center}
\includegraphics[width = 3 in]{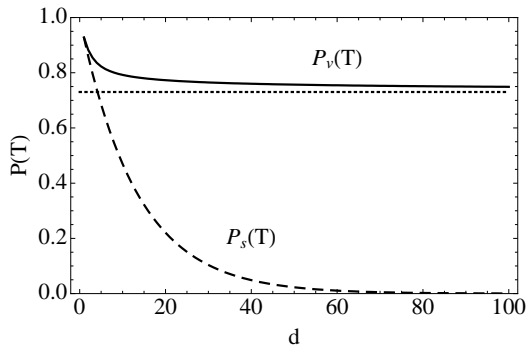}
\caption{The hitting probabilities $P_v(T)$ (solid) and $P_s(T)$ (dashed) as a function of hypercube dimension $d$.  The parameters are $\omega=1$, $\lambda = 1/5$, and $T = \pi/2$.  Also shown is the limiting value $\exp(-\lambda T)$ (dotted) as described in the text.}
\label{qwlimit}
\end{center}
\end{figure}

There are, however,  some important differences between the two models.  First, it is clear that the probability oscillations decay more quickly in the subspace model with $d>1$ than in the vertex model.  Second, while the maximal hitting probability (which occurs near $t=T = \pi/2 \omega = \pi/2$) decreases with increasing $d$ for both models, it does so much more dramatically for the subspace model.  This is further explored in Fig.~\ref{qwlimit}.  As $d$ increases, the probability $P_s(T)$ for the subspace model decays to zero, while $P_v(T)$ for the vertex model converges to a nonzero value for large $d$.  

The behavior of $P_s(T)$ can be rather simply explained.   Using the result of Eq.~(\ref{ssprob}), for small $\lambda$,
\begin{equation}
P_s(T) \approx 2^{-d} \left(1 + e^{-\lambda T/2}\right)^d \approx (1-\lambda T/4)^d \approx e^{- d \lambda T/4}.
\end{equation}
Thus, so long as $\lambda$ and $T$ are independent of $d$, the hitting probability decays exponentially in $d$ for the subspace model.

The behavior of $P_v(T)$ is somewhat more subtle.  First, it is clear from Eq.~(\ref{vlamb}) that $\lambda_{pn} < \lambda$, and thus $e^{-\lambda_{pn} t} > e^{-\lambda t}$.  This can be used in Eqs.~(\ref{vprob1}) and (\ref{vprob2}) to show that
\begin{equation}
P_v(T) = \sum_{n=0}^d g_n(T) > e^{-\lambda T}.
\label{vbound}
\end{equation}
This lower limit is shown as the dotted curve in Fig. \ref{qwlimit}, and is in good agreement with the limiting behavior of the actual $P_v(T)$.  Most importantly, this limit on the hitting probability of the vertex model is independent of $d$.

This striking distinction of the vertex and subspace models requires some discussion.  First, as was noted after Eqs. (\ref{vmodel}) and (\ref{ssmodel}), the subspace model appears to get less information (i.e. about subspaces not vertices) yet causes greater decay of the off-diagonal elements of the density matrix.  This paradox can also be seen in terms of the Lindblad operators, of which the subspace model has $2 d$, while the vertex model has $2^d$.  Presumably more operators should cause more decoherence, but this is not the case.   

In fact, the subspace model involves more measurements than the vertex model.  To understand why this is the case, note that both of these master equations can be derived from the following discrete-time process \cite{Kendon2006}:
\begin{equation}
\rho(t+\delta t) = (1- m p) U \rho(t) U^{\dagger} + p \sum_{j} P_j U \rho(t) U^{\dagger} P_j.
\end{equation}
This non-unitary evolution describes unitary evolution given by $U = e^{-i H \delta t}$ interrupted by $m$ measurements specified by the projectors $P_j$, and each measurement occurring with probability $p = \lambda \delta t$.   The results of these measurements are then ignored, leading to a loss of information.  The projectors here are equal to the Lindblad operators specified above.  The operators for the subspace model can be grouped into $m=d$ measurements (each with 2 projectors summing to the identity), while the vertex model has only one ($m=1$) measurement (with $2^d$ projectors summing to the identity).  Thus, while each measurement in the subspace model extracts less information, it actually performs $d$ measurements, which in the end extracts the same amount of information as the vertex model ($d$ bits).  

As a second and related point, the original work by Alagi\'{c} and Russell used a decoherence rate $\lambda = p/d$ (note that in their notation $d=n$).  Naively, this would remove the exponential decay seen above.  However, they also used an energy scale $\omega = k/d$, such that the relevant parameter $\lambda T = \pi \lambda / (2 \omega) = \pi p / (2 k)$ remains fixed, recovering the exponential decay of $P_s(T)$.  Clearly, allowing the parameters of the model to vary with $d$ can remove the exponential decay; the comparison presented above has the nice feature that in both the vertex and subspace models the system loses information at the same rate ($d \lambda$ bits per second).   

As an alternative perspective, consider a physical hypercube network of qubits designed to implement quantum state transfer \cite{Christandl2004}.  In such a network, one expects that it is the number of operators in the system Hamiltonian that depends on $d$, and not their strength.  The difference between the subspace and vertex models then consists in how the qubits are coupled to an environment (causing decoherence).  As shown in the Appendix, if each qubit has its own environment with an independent dephasing process, the corresponding master equation coincides with the vertex model.  However, it is also shown that if certain sets of qubits share multiple environments, chosen in such a way that qubits in the same subspace of the hypercube undergo a collective dephasing process, the corresponding master equation will describe the subspace model.  Thus, the two decoherence models correspond to different physical systems. 

\begin{figure}
\begin{center}
\includegraphics[width = 3 in]{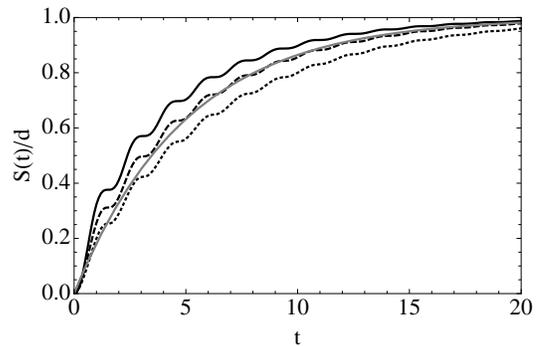}
\caption{The entropy $S(t)$ as a function of time for the quantum walk with decoherence in the vertex model.  The parameters are $\omega=1$, $\lambda = 1/5$, and $d=1$ (solid black), $d=4$ (dashed), and $d=10$ (dotted), and each curve has been scaled by $1/d$.  For comparison the function $1 - e^{-\lambda t}$ is also shown (solid gray).  As described in the text, the entropy for the subspace model is identical to the $d=1$ results (solid black).}
\label{qwentropy}
\end{center}
\end{figure}

The heuristic information arguments given above are confirmed by numerical calculations of the entropy $S(t) = -\text{tr}[\rho(t) \log_2 \rho(t)]$, shown for the vertex model in Fig.~\ref{qwentropy}, which has the qualitative form $S(t) \sim d (1-e^{-\lambda t}) \approx d \lambda t$ for small $t$.  While these calculations are for the vertex model, in the exact solution of the subspace model the total density matrix is equal to the tensor product of $d$ copies of the $d=1$ density matrix.  Thus, the entropy for the $d$-dimensional case is simply $d$ times that for $d=1$, for which the vertex and subspace models are identical.  It is also interesting to observe that the entropy production ($dS/dt$) is greatest when the quantum walk is halfway between the initial and final vertices (e.g. $ t = T/2 = \pi/4$).  At this time the system would, in the absence of decoherence, be in a coherent superposition over the entire hypercube.  Such a state is greatly disturbed by the position (vertex or subspace) measurements, leading to a large increase in entropy.  As seen in the figure, small oscillations in the entropy occur for longer times.  These have twice the frequency of the oscillations in the hitting probability, and occur each time the quantum walk is superposed over the hypercube.

Finally, this comparison of information can be used to reconcile the vertex model with the exponential decay found by Kendon and Tregenna \cite{Kendon2003}.  They observed that the discrete-time hitting probability $P_d$ decayed as $P_d \sim e^{-d p}$, where $p$ is the probability of making a measurement of the position (vertex) after each time step.  Since the number of time steps for this quantum walk to hit the opposing vertex is $\sim d \pi/2$, the total amount of information (with $d$ bits per measurement) extracted during the walk is approximately $d^2 p \pi/2$.   For a true comparison, then, I equate this information to the total information extracted in the two models above ($\approx d \lambda T$ bits), and solve for $p$.  The result, $p \approx 2 \lambda T / (\pi d)$, leads to $P_d \sim e^{- 2 \lambda T/\pi}$, in approximate agreement with Eq. (\ref{vbound}) above.  Thus, so long as one performs less than approximately one complete position measurement for the entire walk, both the discrete and continuous vertex model will hit the opposite vertex with appreciable probability, independent of $d$.  

\section{Conclusion}

In conclusion, I have reconsidered the effects of decoherence in the continuous-time quantum walk on the hypercube.    By considering a particular model of decoherence (the vertex model), I have found that a quantum state initially localized at one vertex will propagate to the opposite vertex with a probability bounded by a constant that is independent of the size of the hypercube.  This is in agreement with an earlier study \cite{Kendon2003} of a discrete-time quantum walk, once the information lost to decoherence has been considered.  This is in striking disagreement with another model of decoherence (the subspace model)  \cite{Alagic2005}, in which the same amount of information is lost, but in an apparently much noisier fashion.  I now consider the interpretation of this result.

The subspace model can be interpreted as extracting information by a sequence of $d$ measurements, one for each dimension of the hypercube, whereas the vertex model uses a single measurement.   By making more measurements to extract the same amount of information, the subspace model disturbs the quantum walk much more than the vertex model.  The number of these measurements is proportional to the hypercube dimension, and each disturbance prevents the walk from reaching its destination.  It is this proliferation of measurements that leads to an exponential decay of the hitting probability as the hypercube gets larger and larger.  

Further understanding this nontrivial relationship between information, measurements, and decoherence in quantum walks may be of some value.   Indeed, the role of decoherence in the mixing of quantum walks remains to be exploited in quantum algorithms \cite{Kendon2006}, and may have implications for biology and beyond \cite{Mohseni2008}. 

\begin{acknowledgments}
I thank Carl Williams at the National Institute of Standards and Technology for early encouragement of this work, and William K. Wootters for helpful discussions.
\end{acknowledgments}

\appendix*
\section{Quantum state transfer}
Quantum state transfer involves a network of qubits interacting with a Hamiltonian of the form
\begin{equation}
\mathcal{H} = \frac{1}{2} \sum_{j < k} \hbar \Omega_{j k} \left( X_j X_k + Y_j Y_k \right)
\label{qbit1}
\end{equation}
where $X$, $Y$, and $Z$ are the Pauli matrices for each qubit, and the coupling matrix is given by  $\Omega_{jk}$.   Decoherence in quantum state transfer can be modeled by a master equation of the general Linblad form
\begin{equation}
\partial_t \rho = -i [\mathcal{H}/\hbar,\rho] + \sum_{j} \lambda_j  (L_j \rho L_j^{\dagger} - \frac{1}{2} L_j^{\dagger} L_j \rho - \frac{1}{2} \rho L_j^{\dagger} L_j ),
\label{qbit2}
\end{equation}
with Lindblad operators $L_j$ and rates $\lambda_j$.  In this Appendix I will illustrate how two different models of decoherence for a hypercube network qubits lead to the quantum walk decoherence models discussed in the text.  

The first is an independent decoherence model, in which the Lindblad terms are given by
\begin{equation}
\sum_{j} T_1^{-1} \left(\sigma^{-}_j \rho \sigma^{+}_j - \frac{1}{2} \{ \sigma^{+}_j \sigma^{-}_j, \rho \} \right)+\frac{1}{2}  \sum_{j} T_{\phi}^{-1} \left(Z_j \rho Z_j - \rho \right),
\label{dmat1}
\end{equation}
with $\sigma^{\pm}_j =(X_j \mp i Y_j)/2$.  This model involves independent energy decay ($T_1$, also called amplitude damping) and dephasing ($T_{\phi}$, also called phase damping) processes for each qubit.  Using this in Eq.~(\ref{qbit2}), and assuming that the initial density matrix $\rho(t=0) = |x)(x|$, where $|x) = X_x |0 \cdots 0\rangle$ ($X_x$ is the Pauli operator for the qubit at location $x$), it will subsequently evolve to
\begin{equation}
\rho(t) = \rho_{0,0}(t) |0) (0| + \sum_{x,y}\rho_{x,y}(t) |x)(y|,
\end{equation}
where $|0) = |0 \cdots 0 \rangle$.  More general general initial states are considered in \cite{Strauch2007}.  These density matrix elements satisfy the differential equations:
\begin{equation}
\begin{array}{lcl}
\partial_t {\rho}_{0,0} &=& T_1^{-1} \sum_{x} \rho_{x,x} = T_1^{-1} (1-\rho_{0,0}) \\
\partial_t {\rho}_{x,y} &=& -i \sum_{z} ( \Omega_{x,z} \rho_{z,y} - \rho_{x,z} \Omega_{z,y}) \\
& &  - T_1^{-1} \delta_{x,y} \rho_{x,x} - 2 [(2 T_1)^{-1} + T_{\phi}^{-1}] \rho_{x,y}(1-\delta_{x,y}).
\end{array}
\label{masteq1}
\end{equation} 
Note that $\rho_{0,0} + \sum_x \rho_{x,x} = 1$, and probability is decaying from the excited-state subspace to the ground state at rate $1/T_1$.  To isolate decoherence within the excited-state subspace I let $\rho_{x,y} = e^{-t/T_1} \tilde{\rho}_{x,y}$, to find that $\tilde{\rho}_{x,y}$ satisfies
\begin{equation}
\partial_t \tilde{\rho}_{x,y} =  -i \sum_{z} ( \Omega_{x,z} \tilde{\rho}_{z,y} - \tilde{\rho}_{x,z} \Omega_{z,y}) + 2  T_{\phi}^{-1} \tilde{\rho}_{x,y}(1-\delta_{x,y}).
\end{equation}
This will agree with the vertex model given by Eq.(\ref{vmodel}) if $H_{j,k} = \Omega_{j,k}$ and $\lambda = 2 T_{\phi}^{-1}$.   The effects of this type of decoherence on the fidelity of perfect state transfer are discussed in \cite{Strauch2007}.

The second model is a collective decoherence model, in which the Lindblad terms are given by
\begin{equation}
\frac{1}{2} \sum_{j} \sum_{\alpha =0}^1T_{\phi}^{-1} \left(S_{j,\alpha} \rho S_{j,\alpha} - \frac{1}{2} \{S_{j,\alpha}^2, \rho \} \right),
\label{dmat2}
\end{equation}
with Lindblad operators
\begin{equation}
S_{j,\alpha} = I + \sum_{\{x | x_j = \alpha\}} (Z_x - I),
\end{equation}
with $I$ the identity operator and $Z_{x}$ the Pauli matrix for the qubit at location $x$.  These operators have been chosen such that $S_{j,\alpha} |x) = - |x)$ if the $j$-th bit in the binary representation of $x$ equals $\alpha$, otherwise $S_{j,\alpha} |x) = + |x)$.  This choice is not unique: $S_{j,\alpha} = \prod_{\{x | x_j = \alpha\}} Z_x$ is another valid choice.  For both cases, using the same initial condition as before, and the result that
\begin{equation}
S_{j,\alpha} |x) (y| S_{j,\alpha} = (2 \delta_{x_j,y_j}-1) |x)(y|,
\end{equation}
I find that $\rho_{0,0}(t) = 0$ and the following differential equation for $\rho_{x,y}(t)$:
\begin{equation}
\begin{array}{lcl}
\partial_t {\rho}_{x,y} &=& -i \sum_{z} ( \Omega_{x,z} \rho_{z,y} - \rho_{x,z} \Omega_{z,y}) \\
& &  - 2 T_{\phi}^{-1} \rho_{x,y} \sum_{j} (1-\delta_{x_j,y_j}).
\end{array}
\label{masteq2}
\end{equation} 
This will agree with the subspace model given by Eq.(\ref{ssmodel}) if $H_{j,k} = \Omega_{j,k}$ and $\lambda = 2 T_{\phi}^{-1}$.

\end{document}